\documentclass[a4paper,11pt]{article}
\usepackage{pos}
\usepackage{textcomp}
\usepackage{enumitem}
\usepackage{graphicx}
\usepackage{dsfont}
\usepackage{bm}
\usepackage{xcolor}
\usepackage{subfig}
\usepackage{caption}
\usepackage{comment}
\usepackage{bbold}
\usepackage{cancel}

\newcommand{\be}{\begin{equation}}
\newcommand{\ee}{\end{equation}}
\newcommand{\beq}{\begin{equation}}
\newcommand{\eeq}{\end{equation}}
\newcommand{\bea}{\begin{eqnarray}}
\newcommand{\eea}{\end{eqnarray}}

\newcommand{\ket}[1]{\ensuremath{| {#1} \rangle }}
\newcommand{\bra}[1]{\ensuremath{\langle {#1} |}}

\newcommand{\beqn}{\begin{eqnarray}}   
\newcommand{\eeqn}{\end{eqnarray}}

\title{Non-Perturbative Bounds for Semileptonic Decays in Lattice QCD}
	
\author[a]{M. Di Carlo}	
\author[b]{G.Martinelli}
\author*[c,e]{M. Naviglio}
\author[d]{S. Simula}
\author[d]{F. Sanfilippo}
\author[e,f]{L. Vittorio}

\affiliation[a]{Higgs Centre for Theoretical Physics, School of Physics and Astronomy, The University of Edinburgh, \\ Edinburgh EH9 3FD, United Kingdom}

\affiliation[b]{Physics Department and INFN Sezione di Roma La Sapienza, \\ Piazzale Aldo Moro 5, 00185 Rome, Italy}

\affiliation[c]{Dipartimento di Fisica dell’Università di Pisa, Largo Bruno Pontecorvo 3, I-56127 Pisa, Italy}

\affiliation[d]{Istituto Nazionale di Fisica Nucleare, Sezione di Roma Tre,\\
Via della Vasca Navale 84, I-00146 Rome, Italy}

\affiliation[e]{Istituto Nazionale di Fisica Nucleare, Sezione di Pisa, Largo Bruno Pontecorvo 3, I-56127 Pisa, Italy}

\affiliation[f]{Scuola Normale Superiore, Piazza dei Cavalieri 7, I-56126, Pisa, Italy}

\emailAdd{matteo.dicarlo@ed.ac.uk}
\emailAdd{guido.martinelli@roma1.infn.it}
\emailAdd{manuel.naviglio@phd.unipi.it}
\emailAdd{silvano.simula@roma3.infn.it}
\emailAdd{francesco.sanfilippo@infn.it}
\emailAdd{ludovico.vittorio@sns.it}

\abstract{We present a new method aiming at a non-perturbative, model-independent determination of the momentum dependence of the form factors entering semileptonic decays using unitarity and analyticity constraints. We extend the original proposal and, using suitable two-point functions computed non-perturbatively, we determine the form factors at low-momentum transfer $q^2$ from those computed explicitly on the lattice at large $q^2$, without making any assumption about their $q^2$ dependence. As a training ground we apply the new method to the analysis of the lattice data of the semileptonic $D \rightarrow K \ell \nu_{\ell}$ decays obtained both at finite values of the lattice spacing and at the physical pion point in the continuum limit. We show that, starting from a limited set of data at large $q^2$, it is possible to determine quite precisely the form factors in a model independent way in the full kinematical range, obtaining a remarkable agreement with the direct calculation of the form factors. This finding opens the possibility to obtain non-perturbatively the form factors entering the semileptonic B decays in the full kinematical range. 
}

\FullConference{%
 The 38th International Symposium on Lattice Field Theory, LATTICE2021
  26th-30th July, 2021
  Zoom/Gather@Massachusetts Institute of Technology
}

\begin{document}
\maketitle

\section{The Dispersive Matrix method}
The Fourier transform of the T-product defines the following Hadronic Vacuum Polarization (HVP) tensors: 
\begin{eqnarray}
\Pi_V^{\mu\nu}(q)&=& i \,\int d^4x \, e^{iq\cdot x}\,  \bra{0} T\{V^{\mu\dagger}(x) V^{\nu}(0)\} \ket{0}\\
 &=& (q^{\mu}q^{\nu}-g^{\mu\nu}q^2) \, \Pi_{1^{-}}(q^2) + q^{\mu}q^{\nu}\, \Pi_{0^{+}}(q^2)\, , \nonumber \\
\Pi_A^{\mu\nu}(q)&=& i \, \int d^4x\,  e^{iq\cdot x} \, \bra{0} T\{A^{\mu\dagger}(x) A^{\nu}(0)\} \ket{0}\label{eq:Pis}\\
 &=& (q^{\mu}q^{\nu}-g^{\mu\nu}q^2)\,  \Pi_{1^{+}}(q^2) + q^{\mu}q^{\nu}\, \Pi_{0^{-}}(q^2)\, , \nonumber 
\end{eqnarray}
where $V^{\mu}=\bar{c}\gamma^\mu b$ , $A^{\mu}=\bar{c}\gamma^\mu \gamma^5 b$ are the weak currents and the subscripts $0^{\pm}$,$1^{\pm}$ represent spin-parity quantum numbers of the intermediate states.\\
The quantities $\Pi_{0^{\pm}}$,$\Pi_{1^{\mp}}$ are called \emph{polarization functions}. In particular, the term proportional to $\Pi_{0^{+}}$ ($\Pi_{0^{-}}$) represents the \emph{longitudinal} part of the HVP tensor with vector (axial) four-currents, while the term proportional to $\Pi_{1^{-}}$ ($\Pi_{1^{+}}$) is the \emph{transverse} contribution to the HVP tensor with vector (axial) four-currents.

\subsection{The starting point}
The imaginary parts of the longitudinal and  transverse polarization functions are related to their derivatives with respect to $q^2$ by the dispersion relations 
\begin{eqnarray}
\chi_{0^{+}} (q^2)&\equiv& \frac{\partial}{\partial q^2} [q^2  \Pi_{0^{+}} (q^2)]=\frac{1}{\pi} \int_0^{\infty} dz \frac{z \,{\rm Im }\Pi_{0^{+}}(z)}{(z-q^2)^2}\, , \nonumber \\[2mm]
\label{veax21}\\[2mm]
\chi_{1^{-}} (q^2)&\equiv& \frac{1}{2} \left( \frac{\partial}{\partial q^2} \right)^2 [q^2  \Pi_{1^{-}} (q^2)]=\frac{1}{\pi} \int_0^{\infty} dz \frac{z \, {\rm Im }\Pi_{1^{-}}(z)}{(z-q^2)^3}\, . \nonumber\\[2mm]
\end{eqnarray} 
In what follows we will denote by $\chi$ a generic susceptibility.  
\\
By inserting a complete set of states with the same quantum numbers of a generic current $J$ we have\footnote{For simplicity we omit Lorentz indices and other complications that are immaterial for the present discussion.}  
\begin{eqnarray}\label{eq: Imaginary}
{\rm Im} \Pi_{0^{\pm},1^{\mp}}=\frac{1}{2} \sum_n \int d\mu(n) (2\pi)^4 \delta^{(4)}(q-p_n) \vert \bra{0} J \ket{n}\vert^2 \, , 
\end{eqnarray}
where $d\mu(n)$ is the measure of the phase space for the set of states $n$. By taking in Eq. \eqref{eq: Imaginary} only the  contribution of the lightest state and using analyticity, we can rewrite the dispersion relations for $\chi(q^2)$ as
\begin{equation}
\label{eq:JQ2z}
\frac{1}{2\pi i } \int_{\vert z\vert =1} \frac{dz}{z}   \vert\phi(z,q^2) f(z)\vert^2 \leq \chi(q^2)\, , 
\end{equation}
where $f(z)$ is the generic form factor and the kinematical functions $\phi(z,q^2)$ for the different form factors entering $B \to D^{(*)}$ decays are defined in Eqs. (40)-(42) of Ref. \cite{Unitarity}.

\subsection{Inner product formalism and the matrix}
Let us introduce an inner product defined as
\beq 
\langle g\vert h\rangle =\frac{1}{2\pi i } \int_{\vert z\vert=1 } \frac{dz}{z}   \bar {g}(z) h(z)\, ,  \label{eq:inpro}
\eeq
where $\bar{g}(z)$ is the complex conjugate of the function $g(z)$. Then, the inequality~(\ref{eq:JQ2z}) can simply be written as
\begin{equation}
\label{eq:JQinpro}
  0\leq \langle\phi f\vert \phi f\rangle\leq \chi(q^2)\, , 
\end{equation}
where we have also used the positivity of the inner product.\\
Following Refs.~\cite{Lellouch0}-\cite{Lellouch1}, we define the function $g_t(z)$ as
\begin{equation}
g_t(z) \equiv \frac{1}{1-\bar{z}(t) z}\, , 
\end{equation}
where $\bar{z}(t)$ is the complex conjugate of the variable $z(t)$ defined in Eq.\,(32) of \cite{Unitarity} and $z$ is the integration variable of Eq. \eqref{eq:inpro}.
It is then straightforward to show that  
\begin{equation}
\label{CI}
\langle g_t|\phi f \rangle  = \phi(z(t),q^2)\, f\left(z(t)\right)\, , \qquad   \langle g_{t_m} | g_{t_l} \rangle  = \frac{1}{1- \bar{z}(t_l) z(t_m)}.
\end{equation}

Let us introduce  the matrix 
\begin{equation}
\label{Der}
\mathbf{M} = \left (
\begin{array}{ccccc}
\langle\phi f | \phi f \rangle  & \langle\phi f | g_t \rangle  & \langle\phi f | g_{t_1} \rangle  &\cdots & \langle\phi f | g_{t_n}\rangle  \\
\langle g_t | \phi f \rangle  & \langle g_t |  g_t \rangle  & \langle  g_t | g_{t_1} \rangle  &\cdots & \langle g_t | g_{t_n}\rangle  \\
\langle g_{t_1} | \phi f \rangle  & \langle g_{t_1} | g_t \rangle  & \langle g_{t_1} | g_{t_1} \rangle  &\cdots & \langle g_{t_1} | g_{t_n}\rangle  \\
\vdots & \vdots & \vdots & \vdots & \vdots \\ 
\langle g_{t_n} | \phi f \rangle  & \langle g_{t_n} | g_t \rangle  & \langle g_{t_n} | g_{t_1} \rangle  &\cdots & \langle g_{t_n} | g_{t_n} \rangle  \\
\end{array}\right )  \, .
\end{equation}
In a numerical simulations of lattice QCD the values ${t_1, \cdots, t_n}$ correspond to  the squared 4-momenta at which the FFs have been computed non-perturbatively and that will be used as inputs for constraining the FFs in regions non accessible to the calculation. Differently, the point $t$ is the unknown point where we want to extract the value of the FFs.

Note that the first matrix element in (\ref{Der}) is the quantity directly related to the susceptibility $\chi(q^2)$  through the dispersion relations.
To be more specific, in the case of $B \to D$ decays, in terms of  the longitudinal and  transverse susceptibilities $\chi_{0^+}(q^2)$ and $\chi_{1^-}(q^2)$ we have that:
\begin{eqnarray}
 && \langle \phi_0 f_0 | \phi_0 f_0 \rangle  \leq \,  \chi_{0^+}(q^2)\, ,  \nonumber  \\
&& \langle \phi_+ f_+ | \phi_+ f_+ \rangle \leq\, \chi_{1^-}(q^2) \, , 
\end{eqnarray}
where $\phi_{0,+}$ are kinematical functions whose definition can be in Eq.s (40)-(42) of Ref. \cite{Unitarity}.

\subsection{The bounds}

The positivity of the inner products~(\ref{CI}) guarantees that the determinant of the matrix~(\ref{Der}) is positive semi-definite, namely
\begin{equation}
\label{detpos}
\det \mathbf{M} \geq 0.
\end{equation}

This condition leads to the following unitarity constraints on the form factor $f(t)$
\begin{equation}
\label{loup}
f_{lo}(t, q^2) \leq  f(t) \leq f_{up}(t, q^2)\, , 
\end{equation}
where 
\begin{equation}
\label{loup2}
f_{lo(up)}(t, q^2) \equiv \frac{-\beta(t, q^2) \mp \sqrt{\Delta_1(t) \, \Delta_2(q^2)}}{\alpha\,  \phi(z(t),q^2)\,}\, .
\end{equation}
The quantities $\alpha, \beta$ and $\gamma$ are determinants of minors of the matrix $M$ defined in Eq. (45) of  \cite{Unitarity}.\\
Note that the positivity of the inner product implies that $\alpha$ and $ \Delta_2$ are $t$-independent, $i.e.$ they are given numbers once the susceptibility $\chi(q^2)$ and the lattice QCD inputs are chosen. On the contrary, $\beta$ and $ \Delta_1$ are $t$-dependent. Moreover, only the quantities $\beta$ and $\Delta_2$ depend on the chosen value of $q^2$. At this point, since $\Delta_1 \geq 0$ by construction, the constraint (\ref{loup}) will be acceptable only when $ \Delta_2 \geq 0$. We stress that the unitarity filter $\Delta_2(q^2) \geq 0$ is $t$-independent, which implies that, when it is not satisfied, no prediction for $f(t)$ is possible at any value of $t$.

Thus, by using a direct lattice measurement of the form factors at the points $t_1, t_2, \dots, t_n$ and the two-point functions of the suitable currents we can constrain the form factors in regions of momenta which for several reasons may not be accessible to lattice simulations.
The application to the case of the semileptonic $D \to K$ decays will be presented in Section~\ref{Sec: Results}.

\section{The novelties of our work}

The DM method allows to reconstruct the interval of the possible values of the form factor in a generic point $t$ in a total model independent way and without any assumption or truncation  starting from, also few, known points and the susceptibilities. With respect to the proposal by L. Lellouch and other previous studies, the main novelties in this work are as follows:
\begin{enumerate}
 \item  We determined non perturbatively all the relevant two-point current correlation functions on the lattice which are fundamental to implement the dispersive bounds (i.e. the susceptibilities $\chi$ that appear in the matrix). We also proposed to reduce discretisation errors of the two-point correlation functions by using a combination of non-perturbative and perturbative subtractions which were found very effective in the past;
 \item  A simpler treatment of the lattice uncertainties with respect to the method proposed in Ref. \cite{Lellouch1}.
 \end{enumerate}
For the first point we refer to the detailed discussion contained in \cite{Unitarity}. We will only discuss here the second novelty. 

\subsection{Statistical uncertainties and Kinematical Constraint}
The machinery discussed in the previous Section allows us to compute the lower/upper bounds of $f_{0(+)}(t)$, once we have chosen our set of input data, i.e.~$\{\chi_{0^+(1^-)}, f_{0(+)}(t_1),\cdots,  f_{0(+)}(t_n)\}$. 
Thus, the input data set is made of $2n+2$ quantities: the $n$ values of the scalar form factor $f_0$, the $n$ values of the vector form factor $f_+$ and the two susceptibilities $\chi_{0^+}$ and  $\chi_{1^-}$. For sake of simplicity we are considering the same number of data points for both the scalar and the vector form factors evaluated at the same series of values $t_i$ ($i = 1, \cdots, n$).
The crucial question is, however, how to propagate the uncertainties related to these quantities into the evaluation of the  FFs $f_{0(+)}(t)$ at a generic value of $t$. 
To answer this question, we propose a method different from the one described in Ref.~\cite{Lellouch1}. 
We start by building up a multivariate Gaussian distribution with mean values and covariance matrix given respectively by $\{ f_{0}(t_1),\cdots,f_{0}(t_n), f_{+}(t_1),\cdots,f_{+}(t_n)\}$ and $\Sigma_{ij} = \rho_{ij} \sigma_i \sigma_j$, where $f_{0(+)}(t_i)$ are the form factors extracted from the three-point functions in our numerical simulation on a given set of gauge field configurations, $\sigma_i$ are the corresponding uncertainties, and $\rho_{ij}$ is their correlation matrix (including also correlations between the two form factors). Note that if we have direct access to the data of the simulations the susceptibilities are properly correlated.

For each jackknife/bootstrap event we consider the $(n+1) \times (n+1)$ matrices $\mathbf{M}^0 $ and $\mathbf{M}^+$ (see Eq.~(\ref{Der})) corresponding to the scalar and vector form factors, respectively.
Since $\Delta_1^0 = \Delta_1^+$ is non-negative by definition, then both $\Delta_2^0$ and  $\Delta_2^+$ should be positive (see Eq.~(\ref{loup2})). Thus, we compute $\Delta_2^{0(+)}$ and verify their signs. If either $\Delta_2^0$ or  $\Delta_2^+$ results to be negative, then the event is eliminated from the sample. From the physical point of view, this step can be read as a consistency check between all the input data, namely the susceptibilities and the FFs for that particular bootstrap. At the end of the procedure, we will be left with $\widetilde{N}_{boot} \le  {N}_{boot}$ events.
\\
At $t=0$ the FFs $f_0$ and $f_+$ are subject to the constraint 
\begin{equation}
\label{KC}
f_0(0) = f_+(0).
\end{equation}
In order to satisfy this condition, in the subset of the $\widetilde{N}_{boot}$ events, we select only the $N_{boot}^* \leq \widetilde{N}_{boot}$ events for which the dispersive bands for $f_0$ and $f_+$ overlap each other at $t=0$.  This corresponds to impose the conditions
\begin{eqnarray}
f_{0,up}(0, q^2) & > & f_{+,lo}(0, q^2) ~ , ~ \nonumber \\
f_{+,up}(0, q^2) & > & f_{0,lo}(0, q^2) ~ ,  ~ 
\label{eq:disegua}
\end{eqnarray}
where $f_{lo,up}(t, q^2)$ were defined in Eq.\,(\ref{loup2}) for a generic form factor $f$.
The above condition selects $N_{boot}^* \leq \widetilde{N}_{boot}$ events.
Following Ref.~\cite{Lellouch1} for each of the $N_{boot}^*$ events we define
\begin{eqnarray}
\label{philo}
f_{lo}^*(0) &=& \max[f_{+,lo}(0),f_{0,lo}(0)]\, ,\nonumber \\
\label{phiup}
f_{up}^*(0)&=& \min[f_{+,up}(0),f_{0,up}(0)]\, , 
\end{eqnarray}
so that, putting $f(0) \equiv f_0(0)=f_+(0)$, one has
\begin{equation}
\label{eq:KCstar}
f_{lo}^*(0) \leq f(0) ~ \leq f_{up}^*(0) ~ . ~
\end{equation}
We now consider the form factor $f(0)$ to be uniformly distributed in the range given by Eq.~(\ref{eq:KCstar}) and we add it to the input data set as a new point at $t_{n+1} = 0$.
To be more precise, for each of $N_{boot}^*$ events we generate $N_0$ values of $f(0)$ with uniform distribution in the range $[f_{lo}^*(0), f_{up}^*(0)]$, obtaining a new sample having $\overline{N}_{boot} = N_{boot}^* \times N_0$ events, each of them satisfying by construction both the unitarity filters $\Delta_2^{0(+)} \geq 0$ and the kinematical constraint~(\ref{KC}).

We then consider two modified $(n+2) \times (n+2)$ matrices, $\mathbf{M}_C^0$ and $\mathbf{M}_C^+$, that have one more row and one more column with respect to matrices $\mathbf{M}^0$ and $\mathbf{M}^+$ and contain the common form factor $f(t_{n+1} = 0)$ and then the information of the kinematical constraint.\\
For any point $t$ at which  we want to predict the allowed dispersive band of the form factor $f(t)$ (which can be either $f_0(t)$ or $f_+(t)$) without  directly computing it in our simulation, we compute the matrix $\mathbf{M}_C$ and using Eq.\,(\ref{loup2}) we get $f_{lo}(t)$ and  $f_{up}(t)$. 
This can be done for each of the $N_0$ events. 
Let us indicate the result of the $k$-th extraction by $f_{lo}^k(t)$ and $f_{up}^k(t)$, respectively.
Then, for each of the $N^*_{boot}$ events the lower and upper bounds $\overline{f}_{lo}(t)$ and $\overline{f}_{up}(t)$ can be defined as  
\begin{eqnarray}
\overline{f}_{lo}(t) & = & \min[f^1_{lo}(t),f^2_{lo}(t),\dots,f^{N_{0}}_{lo}(t)]\, , \nonumber \\
\overline{f}_{up}(t) & = & \max[f^1_{up}(t),f^2_{up}(t),\dots,f^{N_{0}}_{up}(t)] ~ . ~
\end{eqnarray}

\subsection{Recombination of the Bootstrap events}

\begin{figure}[htb!]
\centering
 \includegraphics[scale=0.6]{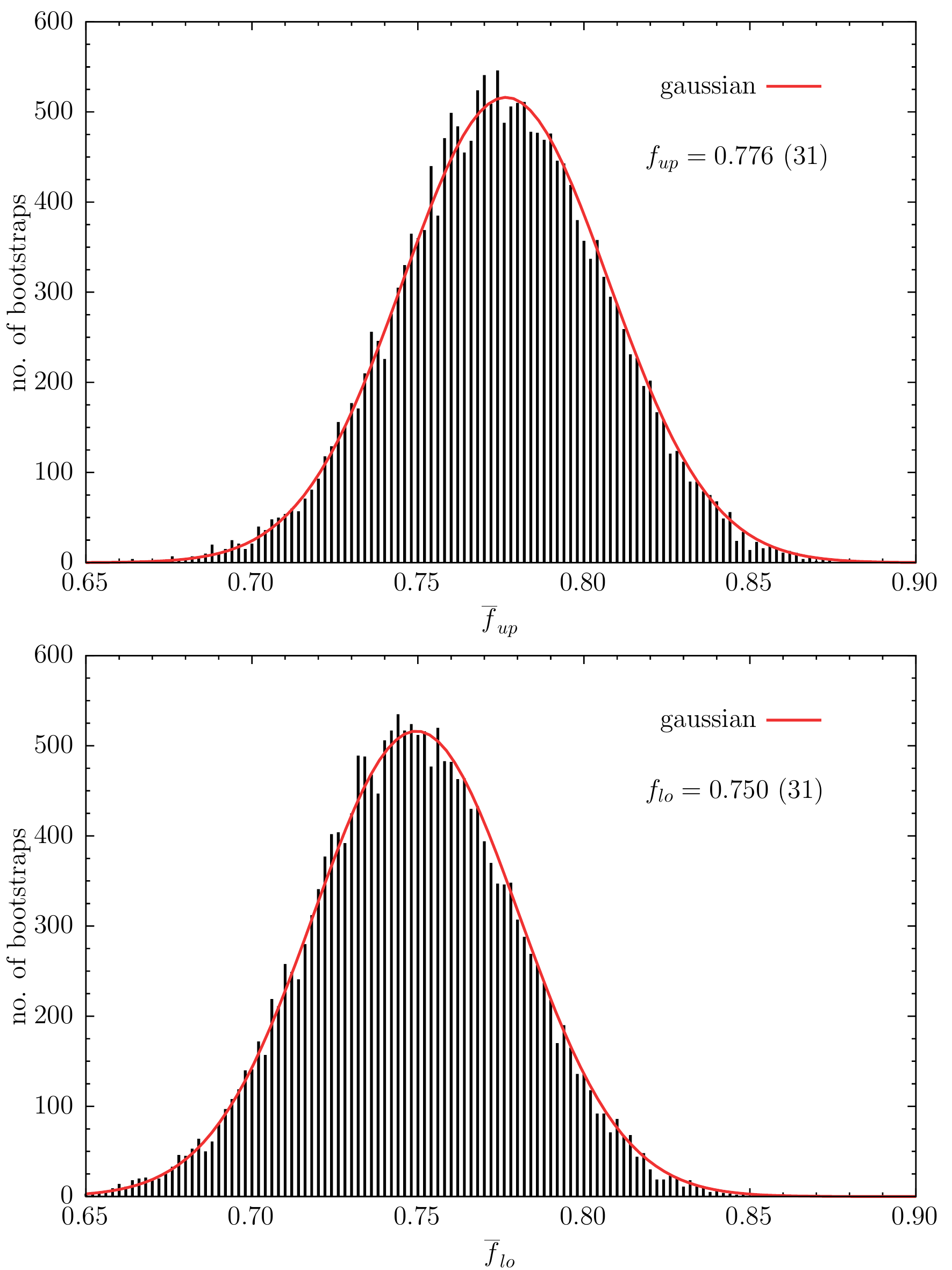}
 \centering
\caption{\textit{Histograms of the values of $\overline{f}_{up}$ (upper panel) and $\overline{f}_{lo}$ (lower panel) for the bootstrap events that pass the unitarity filter in the case of the vector form factor $f_+(t = 0~{\rm GeV}^2)$ of the $D \to K$ transition.}\hspace*{\fill} \small}
\label{Histo_Fit_B}
\end{figure}

At this point we can generate the bounds of the form factor $f(t)$. 
To achieve this goal, we combine all the $N_{boot}^*$ results $\overline{f}_{lo,up}^i(t)$ ($i = 1, \cdots, N_{boot}^*$) to generate the corresponding histograms and fit them with a Gaussian Ansatz, as it is shown in Fig.\,\ref{Histo_Fit_B} in an illustrative case.
From these fits we extract the average values $f_{lo(up)}(t)$, the standard deviations $\sigma_{lo(up)}(t)$ and the corresponding correlation factor $\rho_{lo,up}(t) = \rho_{up,lo}(t)$.
It is understood that the above procedure is applied for both the scalar $f_0(t)$ and the vector $f_+(t)$ form factors.

After these steps, for any choice for $t$ we obtain from the bootstrap events (pseudogaussian) distributions for $f_{0,lo}(t)$,  $f_{0,up}(t)$,  $f_{+,lo}(t)$ and $f_{+,up}(t)$  as well as the corresponding mean values, standard deviations and correlations. 
We combine them according to the procedure described in Sec. V C of \cite{Unitarity} that leads to the final values of the form factor $f(t)$ and its variance $\sigma_f^2(t)$ as
\begin{eqnarray}
f(t) & = & \frac{f_{lo}(t) + f_{up}(t)}{2} ~ , ~ \\
\sigma_f^2(t) & = & \frac{1}{12} \left[ f_{up}(t) - f_{lo}(t) \right]^2 + \frac{1}{3} \left[ \sigma_{lo}^2(t) + \sigma_{up}^2(t) + \rho_{lo,up}(t) \, \sigma_{lo}(t) \, \sigma_{up}(t) \right] ~ . ~
\end{eqnarray}

\section{Results}\label{Sec: Results}
In this Section we discuss two applications of the DM method that show that not only it contains many advantages but it is also very effective in making predictions. In particular, we show that from the knowledge of the form factors in the large $q^2$ region and of the susceptibilities it is possible to determine the form factors with good precision, without making any assumption on their functional dependence on the squared momentum transfer $q^2$.  \\
As an illustration and a test of the method we used the recent results calculation of the form factors in $D \to K$ decays from Ref.\,\cite{Simula}. In the analysis, discussed in paragraph \ref{SubLat}, we make use directly of the results obtained at finite values of the lattice spacing and for unphysical pion masses.\\
We also introduce in paragraph \ref{DavisSec} a further application of the method, not present in Ref. \,\cite{Unitarity} and based on the results of Ref. \,\cite{Davis}.


\subsection{Test of the method in the $D\rightarrow K$ case}\label{SubLat}

Since we have access to the original data of Ref. \cite{Simula}, we can redo the analysis of that paper having computed on the same ensembles also the susceptibilities. The goal is now to implement the matrix method directly on the lattice data points bootstrap by bootstrap and make a totally model-independent extraction of the form factors following the procedure described in the previous section. 

\begin{figure}[htb!]
\centering
\includegraphics[scale=0.70]{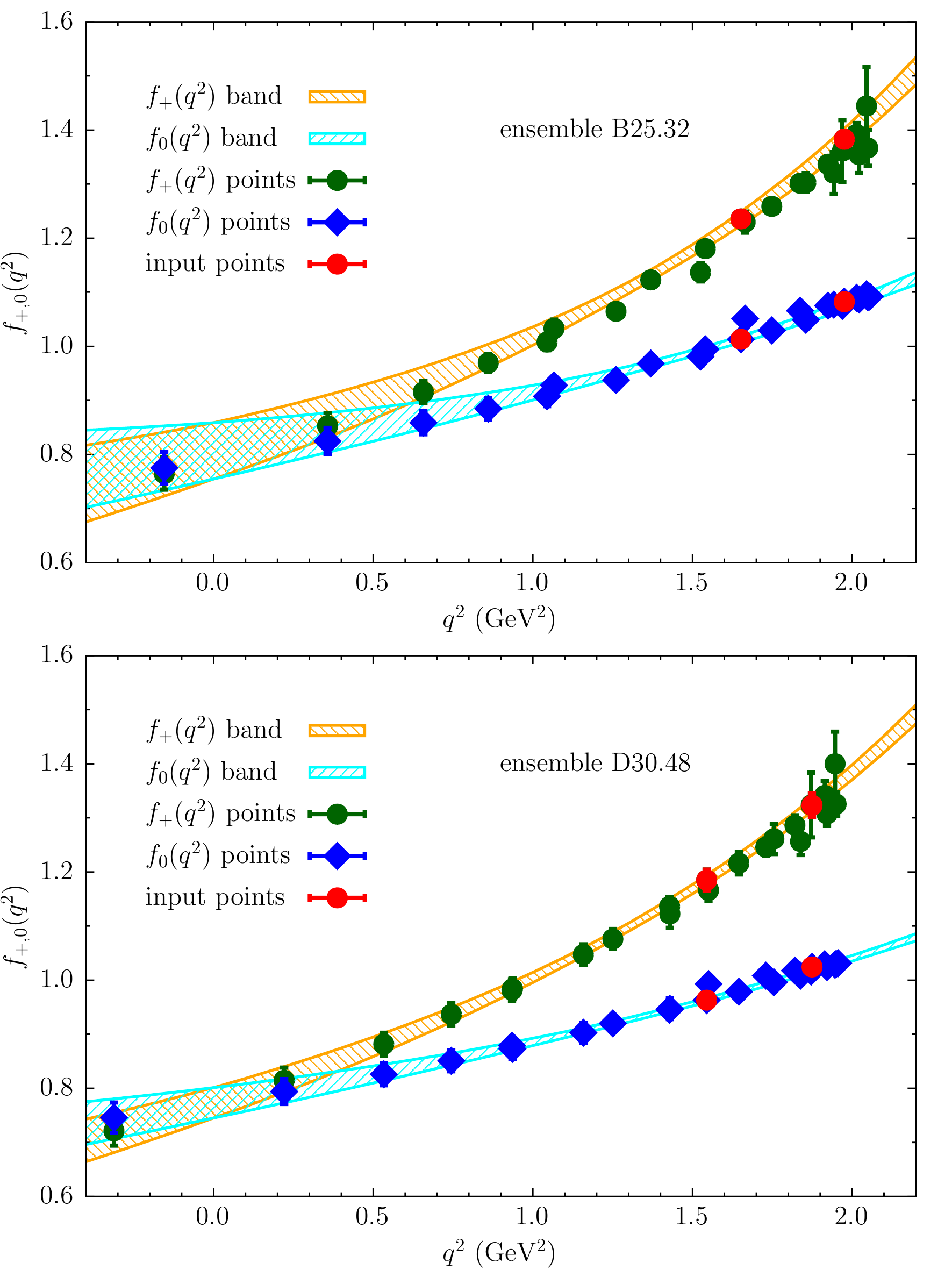}
\caption{\textit{The $D \rightarrow K$ form factors $f_+(q^2)$ (orange band) and $f_0(q^2)$ (cyan band) obtained in this work and in Ref.~\cite{Simula} (dots and diamonds) in the case of the ETMC ensembles B25.32 (upper panel) and D30.48 (lower panel). The red markers (two points at large values of $q^2$ for each form factor) have been used as inputs for our study, while the other ones are not. The lattice data of Ref.~\cite{Simula} are interpolated to the physical values of the charm and strange quark masses determined in Ref.\,\cite{Carrasco}.\hspace*{\fill}}}
\label{Sovrapposizione}
\end{figure}

\begin{figure}[htb!]
\centering
\includegraphics[scale=0.70]{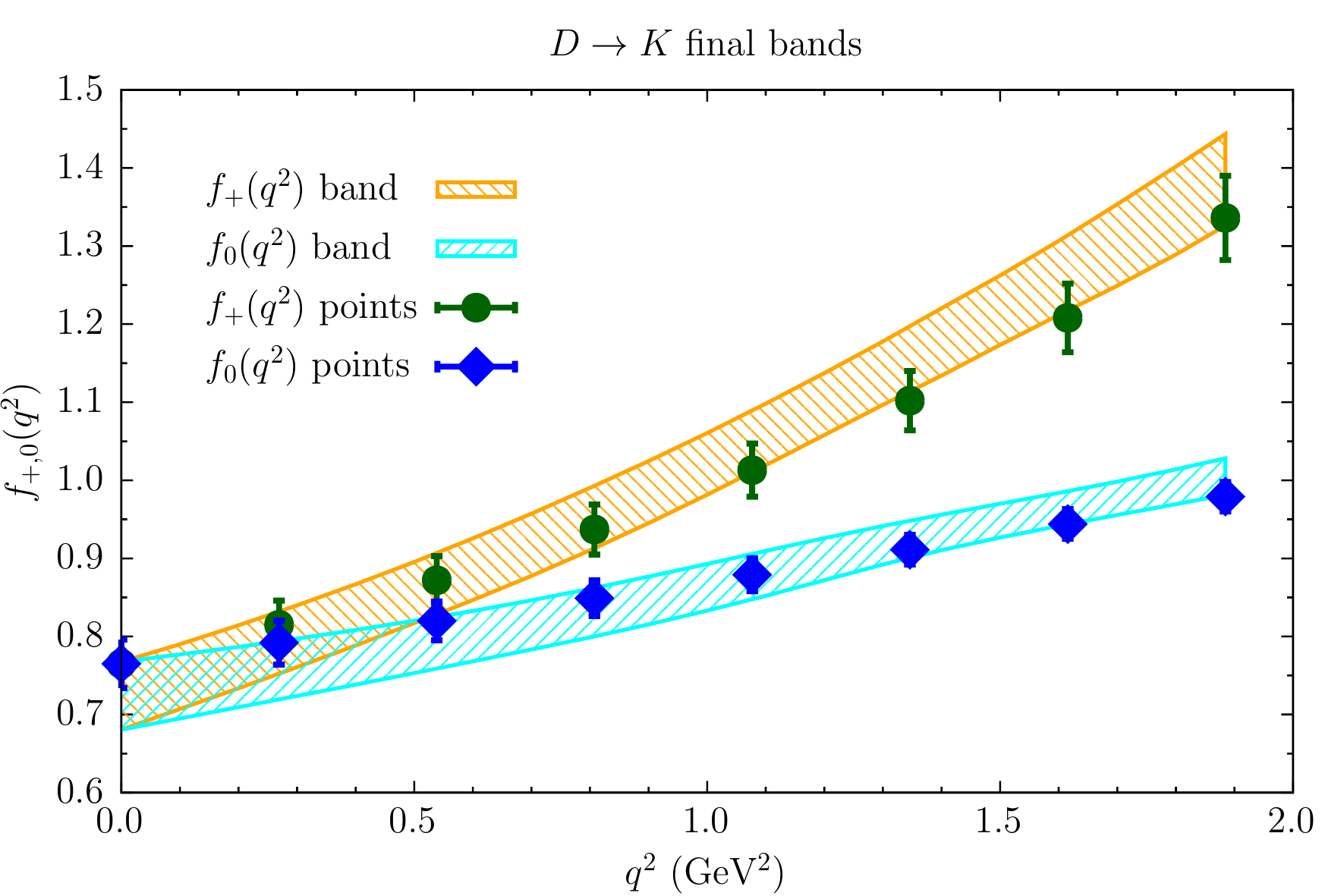}
\centering
\caption{\textit{Momentum dependence of the form factors $f_+(q^2)$ (orange band) and $f_0(q^2)$ (cyan band), extrapolated to the physical point and to the continuum limit, obtained using the dispersive matrix method of this work. The markers represent the lattice results computed in Ref.~\cite{Simula}. \hspace*{\fill}}\small}
\label{Continuo}
\end{figure}

For the $D \rightarrow K$ decay the lattice data of Ref.~\cite{Simula} (already interpolated to the physical charm and strange quark masses) cover all the kinematical region in $q^2$. 
The idea is to mimic what happens in lattice calculations of $B$ decays where all the lattice data are concentrated at $q^2 \sim q^2_{max}$. Thus, we have chosen to use, for each form factor, only two points at large values of $q^2$ corresponding to the $D$-meson at rest, shown as red markers in Fig. \ref{Sovrapposizione}. The great advantage of studying the $D \rightarrow K$ decay is that we can compare our results obtained with the unitarity procedure to the ones obtained from a direct calculation of the form factors. We applied the matrix method described in the previous Sections to the determination of the FFS using 31 bins in $[-0.5 {\rm GeV}^2, q^2_{max}]$. The susceptibilities $\chi_{0^+,1^-}$ are those computed non perturbatively for each ensemble. They have been obtained by eliminating the one particle state both for $\chi_{0^+}$ and $\chi_{1^-}$. Thus, the kinematical functions $\phi_{0(+)}$ have been modified accordingly to Eq.\,(42) of \cite{Unitarity} by including respectively the $D^{*}_s$ and the $D^*_{0}$(2400) poles. Their masses have been calculated on the same ensembles (see Section\,VII B of \cite{Unitarity}). 
To illustrate the procedure we show in Fig.\,\ref{Sovrapposizione} the comparison between our predictions for the allowed bands of the form factors, obtained by using as inputs only the  points denoted as red markers at large $q^2$, and the rest of the lattice points that are not used as input in our analysis in the case of the ETMC ensembles B25.32 and D30.48 (see Appendix~ B of \cite{Unitarity} for details on the simulations). 
The agreement is excellent in all the range. \\
These results suggest that it will be possible to obtain quite precise determinations of the form factors for $B$ decays by combining form factors at large $q^2$ with the non perturbative calculation of the susceptibilities.\\
In Fig.\,\ref{Continuo} we present the final bands for the vector and scalar form factors, extrapolated to the physical value of the pion  mass and to the continuum  limit. The bands agree with the results of Ref.~\cite{Simula} and exhibit a good precision. This demonstrates that the dispersive matrix method allows to determine the semileptonic form factors in their whole kinematical range with a quality comparable to the one obtained by the direct calculations, even if only a quite limited number of input lattice data for each FF (and the non-perturbative susceptibilities) are used\footnote{We have explicitly checked that the results at $q^2 = 0$ shown in Fig.~\ref{Sovrapposizione} are stable against the addition of (red) points provided they are taken in the large $q^2$ region.}.

\subsection{A further application}\label{DavisSec}
Recently the $D \to K$ form factors have been computed quite precisely in \cite{Davis}. In particular, for $q^2=0$ they find
\begin{equation}
f(0) = 0.7380 \pm 0.0043.
\end{equation}
\begin{figure}[htb!]
\centering
\includegraphics[scale=1.0]{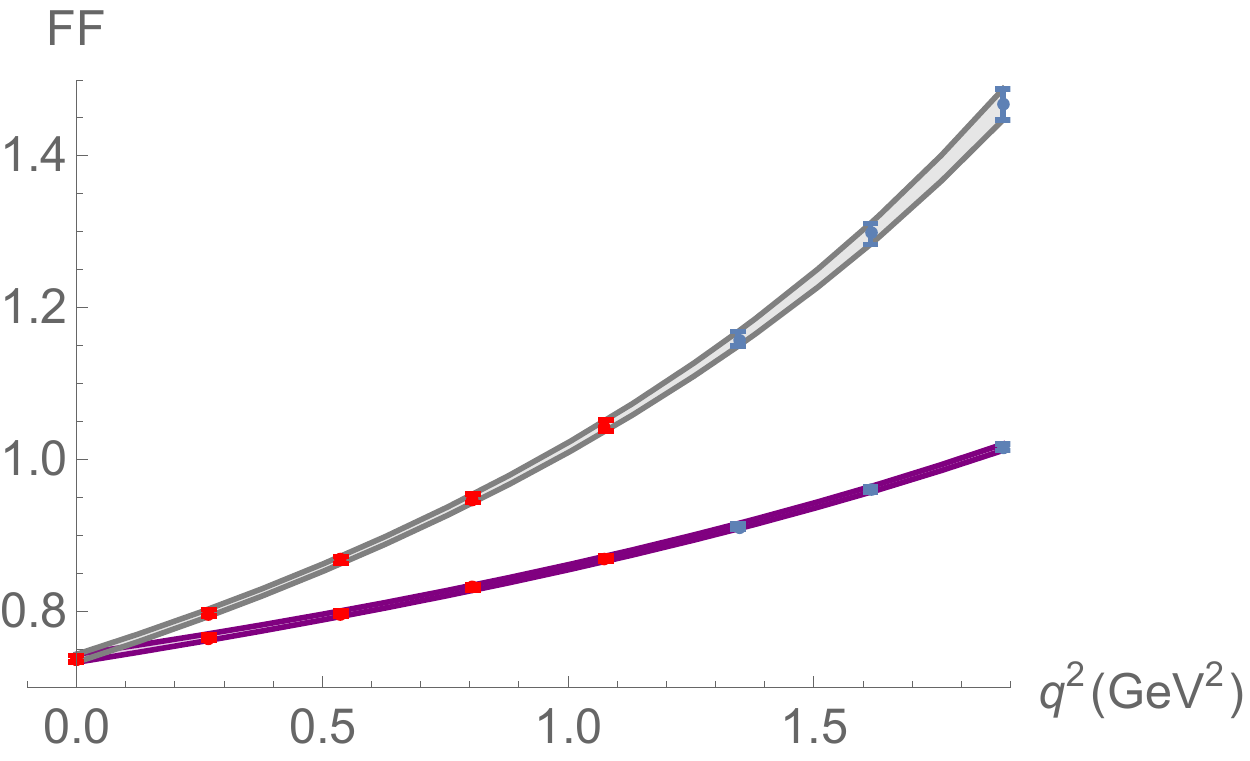}
\centering
\caption{\textit{The $D\rightarrow K$ form factors $f_+(q^2)$ and $f_0(q^2)$ versus the squared 4-momentum transfer $q^2$. The blue and red markers represent a total of eight synthetic data points for each form factor evaluated using the results given in \cite{Davis}.  The grey and purple bands are the results of our dispersive approach obtained using as inputs only the three blue points at the largest values of $q^2$.\hspace*{\fill}}\small}
\label{Dav}
\end{figure}
Using the results given in Table III of \cite{Davis} we have evaluated eight synthetic points for each form factor. Then, we have applied our dispersive analysis using as inputs only the three data at the largest values of $q^2$. The results are shown in Fig. \ref{Dav} in the whole kinematical range. It can be clearly seen that, although we make use of few data points in a limited range of $q^2$, our dispersive bands are remarkably precise and agree with the direct lattice results in the whole kinematical range. In particular, for $q^2=0$ we get
\begin{equation}
f(0) = 0.7384 \pm 0.0052.
\end{equation}
The above findings suggest that, thanks to the DM method, it may be a good strategy to use computing time for improving the precision of few data points at the largest values of $q^2$.

\section{Conclusions and outlook}

We can conclude that the Dispersive Matrix method is very effective and precise in its prediction. It has three main advantages:

\begin{enumerate}
\item The method doesn't rely on any assumption about the functional dependence of the FF on the momentum transferred. Then, it is model independent;
\item It's entirely based on first principles. The susceptibilities and the form factors are non perturbative and we don't rely on any series expansion;
\item It gives very precise and accurate predictions in the whole kinematical range of values of $q^2$ even if we use few data inputs at the largest values of $q^2$.
\end{enumerate}
The DM method has been already successfully applied in \cite{Martinelli:2021myh,Martinelli:2021onb,Martinelli:2021frl} and other applications are underway.

\end{document}